# DESIGN AND OPTIMIZATION OF NANO-BIO-INSPIRED HIERARCHICAL OR NANO-GRAINED MATERIALS

N. Pugno

*Dip. Ingegneria Strutturale, Politecnico di Torino, Torino, Italy; nicola.pugno@polito.it*

**Abstract**
In this letter a mathematical model to design nano-bio-inspired hierarchical materials is proposed. An optimization procedure is also presented. Simple formulas describing the dependence of strength, fracture toughness and stiffness on the considered size-scale are derived, taking into account the toughening biomechanisms. Furthermore, regarding nano-grained materials the optimal grain size is deduced: incidentally, it explains and quantitatively predicts the deviation from the well-known Hall-Petch regime. In contrast with the common credence, this deviation does not arise at a "universal" value of grain size but it is strongly dependent on the mechanical properties of the mixture.

## 1. Introduction

Biological materials exhibit many levels of hierarchy, from the nano- to the macro-scale. For instance, sea shells have 2 or 3 orders of lamellar structures and bone, similarly to dentin, has 7 orders of hierarchy [1]. These nano-bio-materials are composed by hard and strong mineral structures embedded in a soft and tough protein matrix. In bone and dentin, the mineral platelets are ~3nm thick, whereas in shell their thickness is of ~300nm, with very high slenderness. With this hard/soft nano-hierarchical texture, Nature seems to suggest us the key for optimizing materials with respect to both strength and toughness, without losing stiffness. Even if hierarchical materials are recognized to possess a fractal-like topology [2], a mathematical model explicitly considering their complex structure is still absent in literature [3-10]: this represents the aim of our letter. An optimization procedure is also proposed.

A similar optimization procedure based on Quantized Fracture Mechanics [11] is applied to nano-grained materials, thus the optimal grain size is derived. It explains and quantitatively predicts the deviation from the well-known Hall-Petch regime. In contrast with the common credence, this deviation does not arise at a "universal" value of grain size (e.g., around 10nm, as discussed in [12]), but is found to be strongly dependent on the mechanical properties of the mixture.

## 2. The mathematical model

Strength, toughness and stiffness of materials are measured by tensile tests. Imagine a virtual tensile test on a hierarchical bar. Its cross-section, composed by hard inclusions embedded in a soft matrix, is schematized in Figure 1.

The smallest units, at the level $N$, are considered scale-invariant and related to the theoretical material strengths of the hard and soft phases, respectively $\sigma_h$, $\sigma_s$, where usually $\sigma_h >> \sigma_s$. Each inclusion at the level $k+1$ branches into $n_k$ smaller ones, each of them with a surface area $A_k$. Thus, the total number of inclusions at the level $k$ is $N_k = \prod_{j=1}^{k} n_j$.

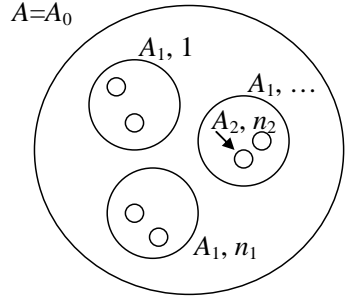

Figure 1: The cross-section of a hierarchical bar.

The equilibrium equation is $F \equiv A\sigma_C = F_h + F_s = N_k A_k \sigma_{hk} + (A - N_k A_k)\sigma_{sk} = N_N A_N \sigma_{hN} + (A - N_N A_N)\sigma_{sN}, \forall k$, where $F$ is the critical applied force, $F_h, F_s$ are the forces carried by the hard and soft phases respectively, $A \equiv A_0$ is the cross-section area of the bar, $\sigma_C$ is its strength, $\sigma_{hN} \equiv \sigma_h$, $\sigma_{sN} \equiv \sigma_s$, and the subscript $k$ refers to the quantities at the level $k$. Note that $\varphi_k = \dfrac{n_{k+1} A_{k+1}}{A_k}$ represents the surface fraction of the inclusions in the matrix, for the level $k$.

Natural optimization suggests self-similar structures [13], for which $n_k = n$ and $\varphi_k = \varphi$, thus $k$-independent; accordingly $N_k = n^k$. Since the inclusions are fractal in nature [14], we expect $F_h \propto R^D$ where $R = \sqrt{A}$ is a characteristic size and $D$ is a constant, the so-called "fractal exponent"; the constant of proportionality can be deduced noting that $F_h(A = A_N) = A_N \sigma_{hN}$, and thus $F = \sigma_{hN} R_N^{2-D} R^D$. Accordingly, from $F_h = \sigma_{hN} R_N^{2-D} R^D = n^N R_N^2 \sigma_{hN}$, we derive:

$$N = D \frac{\ln R/R_N}{\ln n}, \qquad (1)$$

that defines the number of hierarchical levels that we need to design an object of characteristic size $R$. Eq. (1) shows that only few hierarchical levels are required for spanning several orders of magnitude in size. For example, for a nano-structured hierarchical "universe", considering for $R$ its actual radius, i.e., $R \approx 10^{26}$ m, for the smallest units a radius on 1nm, i.e., $R_N \approx 10^{-9}$ m, $n=10$ and $D=1$ would result in only 35 hierarchical levels.

The scaling exponent $D$ can be determined noting that $A - N_N A_N = A(1-\phi)$, where $\phi = \varphi^N$ represents the macroscopic (at level 0) surface fraction of the hard inclusions. Thus, we derive $R/R_N = (n/\varphi)^{N/2}$. Introducing this result into eq. (1) provides the fractal exponent, as a function of well-defined physical quantities:

$$D = \frac{2 \ln n}{\ln n - \ln \varphi}. \qquad (2)$$

Note that $D$ represents the fractal dimension of the inclusions, i.e., of a lacunar surface in which the matrix is considered as empty space; for example for the well-known Sierpinski triangle $D=1.585$. Since $n^N R_N^2 \sigma_{hN} = \sigma_{hN} R_N^{2-D} R^D$ and $R/R_N = (n/\varphi)^{N/2}$, we derive:

$$\phi = \varphi^N = (R/R_N)^{D-2}. \tag{3}$$

Thus, from the equilibrium equation a scaling of the strength is predicted:

$$\sigma_C = \sigma_h \varphi^N + \sigma_s(1-\varphi^N) = \sigma_h(R/R_N)^{D-2} + \sigma_s(1-(R/R_N)^{D-2}). \tag{4}$$

Noting that $n > 1$ and $\varphi < 1$, we deduce $0 < D < 2$ and thus eq. (2) predicts that "smaller is stronger" ($\sigma_h \gg \sigma_s$).

Moreover the *energy balance* implies $W \equiv AG_C = W_h + W_s = A_k G_{hk} + (A - N_k A_k)G_{sk} = N_N A_N G_{hN} + (A - N_N A_N)G_{sN}, \forall k$, where $W$, $W_h, W_s$ are respectively the dissipated energies during fracture in the bar, hard and soft phases, and $G_C$, $G_{hN} \equiv G_h$, $G_{sN} \equiv G_s$ are the fracture energies per unit area of the bar, hard and soft phases respectively; usually $G_h \ll G_s$. Accordingly, the fracture energy must scale as:

$$G_C = G_h \varphi^N + G_s(1-\varphi^N) = G_h(R/R_N)^{D-2} + G_s(1-(R/R_N)^{D-2}). \tag{5}$$

And thus "larger is tougher" ($G_h < G_s$). In the next section, toughening mechanisms will be introduced in the model.

On the other hand, the *compatibility equation* implies (bars in parallels): $K \equiv EA = K_h + K_s = N_k A_k E_{hk} + (A - N_k A_k)E_{sk} = N_N A_N E_{hN} + (A - N_N A_N)E_{sN}, \forall k$, where $K$, $K_h, K_s$ are respectively the stiffnesses per unit length of the bar, hard and soft phases and $E$, $E_{hN} \equiv E_h$, $E_{sN} \equiv E_s$ are the Young moduli of the bar, hard and soft phases respectively. Accordingly, the Young modulus must scale as:

$$E = E_h \varphi^N + E_s(1-\varphi^N) = E_h(R/R_N)^{D-2} + E_s(1-(R/R_N)^{D-2}). \tag{6}$$

Since usually usually $E_h > E_s$, "smaller is stiffer".

Eqs. (4-6) show that at smaller size-scales the inclusions are dominating, whereas at larger size-scales the matrix dominates. These equations present the same self-consistent form: in fact, regarding the generic property $X$ ($\sigma_C$, $G_C$ or $E$) at the non-hierarchical level *N-1*, $X_{N-1} = X_h \varphi + X_s(1-\varphi)$. Thus, at the level *N-2*: $X_{N-2} = X_{N-1}\varphi + X_s(1-\varphi) = X_h \varphi^2 + X_s(1-\varphi^2)$ and iterating $X \equiv X_0 = X_h \varphi^N + X_s(1-\varphi^N)$, as described by eqs. (4-6). In addition, it is clear that the scaling laws predicted by eqs. (4-6) are particularly reasonable, since they predicts two asymptotic behaviour for macro- and nano size-scales. Note that for a three-dimensional architecture (particle inclusions) for which also the third dimension play a role, in the stiffness of eq. (6) the factor of 2 must be replaced by 3, $\varphi$ becomes the volume fraction rather than the surface fraction and *D* is deduced from eq. (2) considering again the factor of 3 instead of 2.

Then, the fracture toughness can be derived as $K_{IC} = \sqrt{G_C E}$, whereas the hardness $H \propto \sigma_C$ formally making the substitution $\sigma_C \to H$ into eq. (4). Note that the important equality (3) would allow us to derive scaling laws from "rules of mixture" also in different systems and for different properties, e.g., for the friction coefficient.

Finally, for quasi-fractal hierarchy, described by $n(R)$ and $\varphi(R)$ weakly varying with the size $R$, a function $D(R)$ must be considered into eqs. (4-6), as deducible by eq. (2).

## 3. Toughening mechanisms: viscoelasticity, plasticity and crack deflection or bridging

Introducing a Young modulus we have not assumed implicitly linear elasticity. For a more realistic behaviour of the matrix, we have to consider visco-elasticity, often observed in bio-tissues. If $\mu_v \equiv (E^0 - E^\infty)/E^\infty$, with $E^0, E^\infty$ short- and long-time elastic moduli respectively ($E^0 \geq E^\infty$, where the equality is valid for linear elasticity), the effective fracture energy becomes $G_s^+ = (1 + \mu_v)G_s$. The parameter $\mu_v$ represents an enhancement factor for fracture energy dissipation due to the viscoelastic properties of the medium, e.g., for bone $\mu_v \approx 4$ or for shell $\mu_v \approx 1.5$ (see [1]). Including plasticity, if $\mu_p$ represents the enhancement factor due to the plastic work during fracture $G_s^+ = (1 + \mu)G_s$, where $\mu = \mu_v + \mu_p$. The factor $\mu_p$ can be estimated noting that the effect of plasticity on the crack propagation is described by a blunting of the crack tip due to dislocation emissions. If $a$ is the "fracture quantum", a material/structural parameter, $\mu_p = \rho/(2a)$ [11], with $\rho$ tip radius.

According to the previous analysis and Fig. 1, the fracture surface is assumed to be planar. On the other hand, the inclusions could serve as hard structures to deflect the crack path or as crack bridging fibers. In the latter case, the inclusions will be pulled-out after fracture, incrementing the dissipated energy in a fashion similar to the former mechanism (if the fracture of the matrix is assumed as that of the interface; it is evident that this hypothesis can be easily removed). To model such effects we simply assume the two-dimensional scheme reported in Fig. 2, a lateral view of the crack surface of Fig. 1.

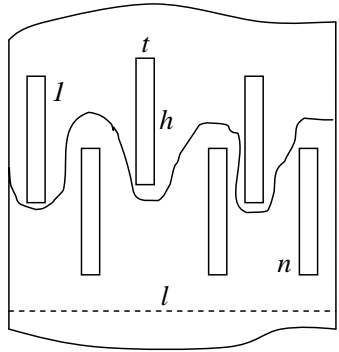

Figure 2: The lateral view of the crack surface.

According to this scheme $G_s^{++} l = G_s^+ (l + nh)$, where $l$ is the nominal crack length, $n$ is the number of inclusions along $l$ and $h$ is their height. Noting that $\varphi = nt/l$ with $t$ thickness of the inclusions and that $\lambda = h/t$ is their slenderness, the effective fracture energy becomes $G_s^{++} = (1 + \lambda\varphi)G_s^+$. Thus, for this toughening mechanism the fracture energy $G_s^{++}$ can be even strongly larger ($\mu \approx \varphi \approx 1; \lambda \gg 1$) than the intrinsic fracture energy of the matrix $G_s$, if large values for $\lambda$ are considered. It explains why the shape of mineral crystals is found to be very anisometric (platelets): the anisometry is larger for bone and dentin (platelets 3nm thick and up to 100nm long) and enamel (15-20nm thick, 1000nm long) than for nacre (i.e., see shells, 200-500nm thick and 5-8μm long) [1]. For details on the hierarchical bone structure see [15]. Thus, eq. (5) has in general to be considered with the substitutions:

$$G_s \rightarrow (1+\mu)(1+\lambda\varphi)G_s \qquad (7)$$

$$G_h \rightarrow 0 \qquad (8)$$

since in this case no dissipations occurs in the hard phase.

Furthermore, a soft matrix activates shear mechanisms rather than longitudinal ones, according to the tension-shear chain model recently proposed [1]. Since in this case matrix does not carry tensile load, the substitution:

$$\sigma_s \rightarrow 0 \qquad (9)$$

must be considered into eq. (4). Considering a linear variation of the shear stress (stress concentration factors could be included on the basis of the analysis reported in [16]) with a maximum value $\tau$ implies a maximum normal stress $\sigma$ in the platelet equal to $\lambda\tau$ [1]. Thus load transfer requires $\lambda\tau_s > \sigma_h$ where $\tau_s$ is the shear strength of the matrix; this shows that low values of $\tau_s$ are compensated by Nature by high slendernesses $\lambda$ [1]. Note that according to [1] an in-serie tension/shear rather than an in-parallel tension architecture, as considered in eq. (6), emerges. However their asymptotic behaviours (for realistic sufficiently large size-scales $R$) are identical if in eq. (6) the Young modulus of the matrix is assumed to be negligible, i.e.:

$$E_s \rightarrow 0 \qquad (10)$$

**4. Hierarchical optimization**

Let us assume for the sake of simplicity the Griffith's case. According to Quantized Fracture Mechanics (QFM) [11] the failure stress is predicted to be $\sigma_f = \sqrt{G_C E/(\pi l + a/2)}$, where $2l$ is the crack length and $a$ is the fracture quantum (linear elastic fracture mechanics assumes $a=0$). Thus, a "flaw tolerance" is expected to take place for crack length $l$ smaller than $a$ and surely in platelets with thickness $t \approx a$. The fracture quantum $a$ can be estimated noting that $\sigma_f(l=0) = \sigma_C$ and thus the optimization against flaw tolerance implies:

$$t_{opt} \approx \frac{G_C E}{\sigma_C^2} \qquad (10)$$

at all the hierarchical levels, similarly to what deduced in [17,18]. In addition, note that Nature seems to optimize the ratio between fractal and Euclidean nominal dimensions towards the intermediate value of ¾ [13]. Thus, from eq. (2), $D/2 \approx 3/4$ and the optimization would also imply:

$$D_{opt} \approx 3/2, \quad \text{or,} \quad (\varphi n^{1/3})_{opt} \approx 1 \qquad (11)$$

The fractal dimension of the inclusions, according to eq. (13), is intermediate between an Euclidean line and surface.

In addition, to reach the failure simultaneously in the soft and hard phases:

$$\lambda_{opt} \approx \sigma_h/\tau_s \qquad (12)$$

## 5. The optimal grain size and the deviation from the Hall-Petch regime

Referring to the Griffith's case according to QFM [11] $\sigma_f = \sqrt{G_C E/(\pi l + a/2)}$; imagine a matrix embedding (not hierarchical) grains having a mean diameter $d$. The characteristic crack length is statistically expected to be of the order of $2l \approx d$ [19]; thus $\sigma_f \approx \sqrt{2G_C E/(\pi d + a)}$. Accordingly, for large grain size $\sigma_f \propto d^{-1/2}$ as described by the well-known Hall-Petch law, whereas for small grain size a deviation from this law is expected, i.e., $\sigma_f \propto d^0$. This deviation is observed in experiments and is erroneously assumed to take place around a grain size of ~10nm (as discussed in [12]). The same transition has been derived in [20], extending the fractal approach for size-effects reported in [14]. The grain size corresponding to the transition, or even to the inversion of the Hall-Petch law (in the case of a positive exponent, for example due to a fracture energy $G_C(d)$ increasing with the grain size $d$) must thus represent an optimal value (for flaw tolerance). According to our simple considerations, and in contrast to the common assumption [12] of a "universal" value of ~10nm, we expect the optimal grain size dependent on the material properties of the mixture but assumed to be very fine grained (i.e., $d \rightarrow 0$, since $\sigma_C = \sigma_f(d=0)$ ) and containing the given grain content $\phi$ (e.g., in the form suggested by eqs. (4-6) with $\varphi^N = \phi$), i.e.:

$$d_{opt} \approx \frac{G_C(\phi)E(\phi)}{\sigma_C^2(\phi)} \approx \frac{K_{IC}^2(\phi)}{\sigma_C^2(\phi)} \tag{13}$$

Note that for $K_{IC} \approx 10\,\text{MPa}$, if $\sigma_C \approx 100\,\text{GPa}$ $d_{opt} \approx 10\,\text{nm}$, whereas if $\sigma_C \approx 1\,\text{GPa}$ $d_{opt} \approx 100\,\mu\text{m}$. We have found some experiments in literature that show a deviation from the Hall-Petch law for a "large" grain size, i.e. of several microns [19, 21], supporting our argument.

Clearly for hierarchical grains embedded in a nano-grained matrix, the two approaches can be mixed by applying rule of mixtures (e.g., in the form suggested by eqs. (4-6)).

Note that the optimal length reported in eq- (13) could perhaps be applied also in different contexts, e.g., to optimizing coating layers.

## 6. Conclusions

In conclusion, the developed mathematical model allow us to preliminary design and optimize nano-bio-inspired hierarchical materials, starting from the theoretical nanoscale level, by following a bottom-up procedure. An grain-size optimum value for grained mixtures is also proposed. No free-parameters are present in the model. Obviously the complexity of the problem has imposed a simplified treatment with associated limitations.